\documentclass[conference]{IEEEtran}
\IEEEoverridecommandlockouts
\usepackage{cite}
\usepackage{amsmath,amssymb,amsfonts}
\usepackage{algorithmic}
\usepackage{graphicx}
\usepackage{textcomp}
\usepackage{xcolor}

\usepackage[ruled,lined]{algorithm2e}

\usepackage{tikz}
\usepackage[most]{tcolorbox}
\usepackage{lmodern}
\usepackage{inconsolata} % Monospaced font
\usepackage{booktabs}
\usepackage{url}
\usepackage{tabularx}

\def\BibTeX{{\rm B\kern-.05em{\sc i\kern-.025em b}\kern-.08em
    T\kern-.1667em\lower.7ex\hbox{E}\kern-.125emX}}

\DeclareRobustCommand*{\IEEEauthorrefmark}[1]{%
    \raisebox{0pt}[0pt][0pt]{\textsuperscript{\footnotesize\ensuremath{#1}}}}

\usepackage{xurl}

\begin{document}

\title{PhishDebate: An LLM-Based Multi-Agent Framework for Phishing Website Detection}

\author{
    \IEEEauthorblockN{WENHAO LI\IEEEauthorrefmark{1}, SELVAKUMAR MANICKAM\IEEEauthorrefmark{1,*}\thanks{SELVAKUMAR MANICKAM is the corresponding author.}, YUNG-WEY CHONG\IEEEauthorrefmark{2}, SHANKAR KARUPPAYAH\IEEEauthorrefmark{1}}
    \IEEEauthorblockA{\IEEEauthorrefmark{1} Cybersecurity Research Centre, Universiti Sains Malaysia, Pulau Pinang, Malaysia}
    \IEEEauthorblockA{\IEEEauthorrefmark{2} School of Computer Sciences, Universiti Sains Malaysia, Pulau Pinang, Malaysia}
    \IEEEauthorblockA{wenhaoli@ieee.org, \{selva, chong, kshankar\}@usm.my}
}

\maketitle

\begin{abstract}
Phishing websites remain a major cybersecurity threat, exploiting deceptive structures, brand impersonation, and social engineering to evade detection. Recent advances in large language models (LLMs) have improved phishing detection through contextual understanding, yet most existing approaches rely on single-agent classification, which is prone to hallucination and often lacks interpretability and robustness. To address these limitations, we propose PhishDebate, a modular multi-agent LLM-based debate framework for phishing website detection. Four specialized agents independently analyze webpage aspects, including URL structure, HTML composition, semantic content, and brand impersonation, under the coordination of a Moderator and final Judge. Through structured debate and divergent reasoning, the framework achieves more accurate and interpretable decisions. By reducing uncertain predictions and providing transparent reasoning, PhishDebate functions as an analyst-augmentation system that lowers cognitive load and supports early, left-of-exploit detection of phishing threats. Evaluations on commercial LLMs show that PhishDebate achieves 98.2\% recall on a real-world phishing dataset and outperforms single-agent and Chain-of-Thought (CoT) baselines. Its modular design enables agent-level configurability, allowing adaptation to varying resource and application requirements, and offers scalability to high-velocity, large-scale security data environments.
\end{abstract}


\section{Introduction}

Phishing attacks continue to pose a pervasive and evolving threat across the digital landscape, targeting individuals, businesses, and institutions alike \cite{10788671}. By mimicking legitimate websites and exploiting human trust, phishing websites deceive users into disclosing sensitive information such as login credentials, financial details, or personal data \cite{10.1145/1073001.1073009,10.1145/3589334.3645535}. The sophistication and scale of phishing operations have escalated in recent years, often leveraging automation, obfuscation, and brand impersonation to evade detection systems \cite{10175532}. Consequently, phishing remains one of the most pressing cybersecurity challenges, accounting for a significant proportion of online fraud and data breaches \cite{10584537}.

To combat this threat, a wide range of detection techniques have been proposed. Traditional approaches relied on heuristic rules and blacklists, which are lightweight but prone to obsolescence in the face of adaptive attackers \cite{5931389,Mohammad2014}. Machine learning models have since emerged, incorporating engineered features from Uniform Resource Locators (URLs), HyperText Markup Language (HTML) structures, and third-party metadata \cite{SAHINGOZ2019345,make3030034}. More recently, deep learning and transformer-based architectures have demonstrated promising results by learning patterns from raw inputs such as URLs, webpage content, and screenshots \cite{9094653,10.1145/3372297.3417233,10.1007/978-3-031-24985-3_28}. With the rise of Large Language Models (LLMs), phishing detection has entered a new era, where contextual understanding and semantic reasoning allow for the identification of subtle social engineering tactics \cite{Gallagher2024}.

However, most existing LLM-based methods for phishing website detection rely on single-agent classification \cite{10723311}, which not only lacks iterative reasoning and interpretability but also risks hallucinations \cite{lin2024interpreting} and narrow decision-making due to dependence on a single perspective. This limitation reduces their robustness and reliability when confronting complex, ambiguous, or deceptive phishing tactics.

To address these limitations, we propose PhishDebate, a modular multi-agent LLM-based debate framework for phishing website detection. PhishDebate comprises four specialized agents, each targeting a distinct dimension of phishing evidence, including URL structure, HTML composition, semantic content, and brand impersonation. A Moderator oversees the structured debate process, while a final Judge delivers the classification verdict based on all arguments. The framework follows a modular design that allows users to flexibly include or exclude agents based on resource constraints or task-specific requirements. This structure supports collaborative reasoning, improves detection accuracy, enhances interpretability, and reduces dependence on a single-agent perspective. Moreover, by reducing uncertain predictions and providing transparent reasoning, PhishDebate acts as an analyst-augmentation system that helps lower the cognitive burden of security analysts. Its ability to identify phishing attempts at an early, left-of-exploit stage underscores its relevance for proactive defense, while the modular design offers scalability to high-velocity, large-scale security data environments.

The key contributions of this paper are as follows:
\begin{itemize}
    \item We introduce PhishDebate, one of the first debate-based multi-agent LLM frameworks specifically designed for phishing website detection, emphasizing modularity, explainability, and collaborative inference.
    \item We develop a set of specialized agent roles and debate coordination strategies that support multi-round argumentation and dynamic consensus evaluation, enabling the detection of complex phishing attempts.
    \item We conduct comprehensive evaluations using two real-world phishing datasets and four state-of-the-art LLMs (e.g., GPT-4o, Gemini-2.0), comparing PhishDebate against baselines including direct prompting and Chain of Thought (CoT) reasoning.
    \item We demonstrate through scenario analysis and case studies that PhishDebate not only improves classification performance but also provides interpretable rationales, enhancing trust and transparency in phishing detection.
\end{itemize}

The remainder of this paper is organized as follows: Section~\ref{sec:related} reviews relevant literature in phishing detection and multi-agent LLM reasoning. Section~\ref{sec:framework} details the design of the PhishDebate framework, including agent roles and debate logic. Section~\ref{sec:eval} presents the evaluation setup, datasets, and performance metrics. Section~\ref{sec:results} reports experimental results, including comparative baselines and scenario analyses and a detailed case study. Finally, Section~\ref{sec:limitations} discusses the limitations and Section~\ref{sec:conclusion} concludes the paper with future directions.

\section{Related Work}\label{sec:related}
\subsection{Phishing Website Detection}

Phishing website detection has evolved significantly over the past few decades, beginning with heuristic-based approaches \cite{5931389,Mohammad2014} that relied on rule-based inspection of URLs, domain features, and HTML signatures. These methods typically used predefined rules such as the presence of IP-based URLs, suspicious JavaScript functions, or invisible elements to flag malicious content. While lightweight and interpretable, heuristic techniques suffered from low adaptability to novel phishing tactics \cite{10788671}.

To improve generalization, researchers adopted machine learning-based classifiers~\cite{SAHINGOZ2019345,make3030034}, which extract engineered features from URLs, HTML structure, and third-party metadata (e.g., WHOIS, Alexa rank) \cite{6731669}. Algorithms like decision trees, support vector machines (SVM), and random forests were trained on labeled datasets to detect phishing behavior patterns. However, their reliance on manual feature extraction and third-party data posed scalability and robustness challenges.

Deep learning methods, especially Convolutional Neural Networks (CNNs) and Recurrent Neural Networks (RNNs), further advanced detection capabilities by automatically learning patterns from raw input such as URLs, text, and screenshots~\cite{9094653,10.1145/3372297.3417233,10.1007/978-3-031-24985-3_28}. Vision-based models using webpage screenshots and hybrid models combining URL, Document Object Model (DOM), and content-level features have shown promise in detecting visually deceptive phishing websites~\cite{272200,279900}. Recently, transformer-based architectures and pre-trained language models have been applied to capture semantic cues in webpage content and URLs~\cite{9653028,10788671}, improving the understanding of social engineering cues.

With the rise of LLMs, phishing website detection has entered a new paradigm~\cite{10788671,10723311}. LLMs provide context-aware understanding of webpage elements, brand impersonation, and social engineering tactics, which enhances detection performance even in the presence of various modern cloaking techniques~\cite{10175532,lisecrypt23}. However, most existing approaches rely on binary classification with single-shot LLM prompts \cite{10723311}, lacking collaborative reasoning or iterative verification. This gap highlights the opportunity for more interpretable, resilient, and robust LLM-based detection frameworks.

\subsection{Multi-Agent Debate and Collaborative Reasoning}

Multi-agent debate systems are inspired by human deliberation, where multiple independent agents analyze and critique a shared problem before reaching a decision \cite{guo2024large}. These systems have been increasingly used to enhance the reasoning abilities of language models by encouraging diversity of perspectives and iterative refinement~\cite{li2023camel,chan2023chateval,liang2024}. Each agent typically specializes in a different domain or analytic skill and engages in structured argumentation rounds moderated by a central judge or aggregator.

In the domain of phishing email detection, a debate-driven multi-agent LLM framework is proposed~\cite{11012014} where two agents argue for and against a phishing classification, and a third judge agent determines the final verdict. This setup enabled the system to uncover subtle social engineering cues and improved detection accuracy without relying on handcrafted features or complex prompt engineering. Similarly, MultiPhishGuard~\cite{xue2025multiphishguard} introduced a five-agent system combining text, URL, metadata, explanation, and adversarial agents, coordinated by reinforcement learning to adapt to evolving adversarial threats. The model achieved good performance with enhanced robustness and interpretability through its explanation simplifier component.

Beyond phishing, the multi-agent debate framework has been applied to spam detection. A recent study~\cite{10868417} employed LLM agents with distinct roles to simulate complex evaluation processes similar to human decision-making. This collaborative design enhanced both accuracy and resilience against evolving spam tactics, outperforming traditional single-agent detectors in precision and adaptability.

These works demonstrate the effectiveness of multi-agent debate systems in security-relevant classification tasks. However, their applications have been primarily limited to phishing emails and spam detection, leaving phishing website detection as an unexplored frontier. This motivates our work to extend the debate paradigm to this domain.

\section{PhishDebate Framework}\label{sec:framework}

In this section, we present PhishDebate, a novel multi-agent LLM framework for phishing website detection. Our approach leverages the collaborative intelligence of specialized agents through structured debate to achieve robust and interpretable phishing detection.

\subsection{Overall PhishDebate Framework}

\begin{figure*}[!t]
\centering
\includegraphics[width=0.95\textwidth]{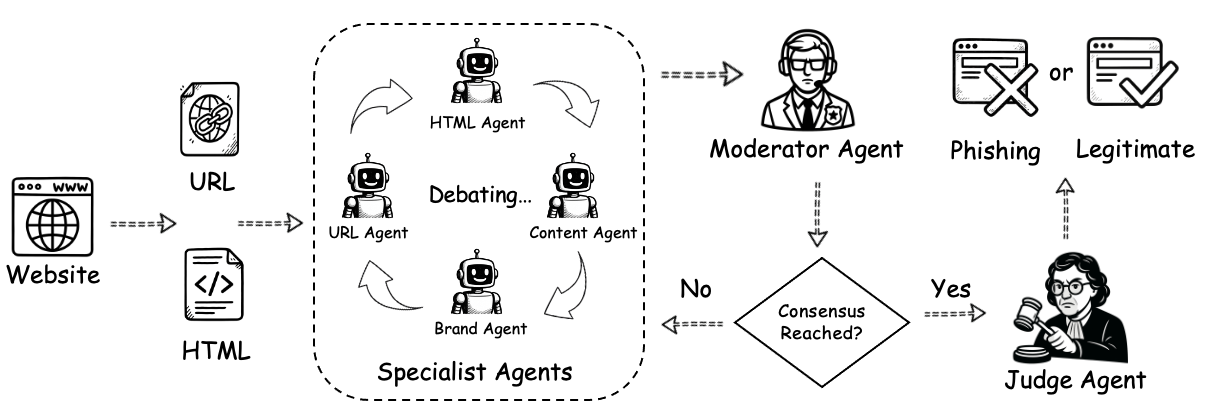}
\caption{The Workflow of PhishDebate Framework.}
\label{fig:workflow}
\end{figure*}

PhishDebate employs a multi-agent debate system consisting of six distinct agents: four specialist agents (URL Analyst Agent, HTML Structure Agent, Content Semantic Agent, and Brand Impersonation Agent), one Moderator, and one Judge. The framework follows a structured debate process where specialist agents analyze different aspects of a website and engage in collaborative reasoning to reach a consensus. The Fig.~\ref{fig:workflow} illustrates the workflow of PhishDebate.

The framework operates through four main phases:
\begin{enumerate}
    \item \textbf{Initial Analysis Phase (Round 1)}, where each active specialist agent independently analyzes the website from their domain expertise without seeing other agents' responses.
    \item \textbf{Consensus Evaluation Phase}, where the Moderator evaluates if consensus exists after each round based on the discussion from specialist agents. If a consensus is reached, no further round will go then.
    \item \textbf{Multi-round Debate Phase (Rounds 2+)}, where agents can see and respond to other agents' analyses, refining their assessments through structured discussion. This phase continues until consensus is reached or maximum rounds are completed.
    \item \textbf{Final Judgment Phase}, where the Judge always makes the final determination based on all available evidence from the debate process, regardless of whether consensus was reached.
\end{enumerate}

Note that if \texttt{$R_{max}$=1}, the system performs only the initial independent analysis followed by consensus evaluation, then proceeds directly to the Judge for final decision-making.

The key innovation of PhishDebate lies in its ability to combine diverse analytical perspectives through structured argumentation, enabling the system to capture subtle phishing indicators that might be missed by single-agent approaches. The debate mechanism ensures that each agent's specialized analysis contributes to the final decision while maintaining transparency through detailed reasoning chains. Algorithm~\ref{alg:phishdebate} presents the proposed PhishDebate framework.

\begin{algorithm}
\caption{PhishDebate Framework}
\label{alg:phishdebate}
\footnotesize
\KwIn{URL $U$, HTML $H$, text $T$, $R_{max}$, $\tau$, agents $A$}
\KwOut{Assessment $F$, confidence, rounds, early termination}

\SetKwFunction{PD}{PhishDebate}
\SetKwFunction{GSP}{GenPrompt}
\SetKwFunction{QA}{QueryAgent}
\SetKwFunction{ME}{ModEval}
\SetKwFunction{AR}{AggResp}
\SetKwFunction{GDP}{GenDebPrompt}
\SetKwFunction{GJP}{GenJudgePrompt}
\SetKwFunction{QJ}{QueryJudge}

\SetKwProg{Fn}{Function}{:}{}
\Fn{\PD{$U, H, T, R_{max}, \tau, A$}}{
    $responses \leftarrow \emptyset$, $round \leftarrow 1$\;
    
    \tcp{Phase 1: Independent Analysis (Round 1)}
    \ForEach{$a_i \in A$}{
        $prompt_i \leftarrow$ \GSP{$a_i, U, H, T$}\;
        $responses[a_i] \leftarrow$ \QA{$a_i, prompt_i$}\;
    }
    
    \tcp{Phase 2: Consensus Evaluation}
    $consensus \leftarrow$ \ME{$responses, round$}\;
    \If{$consensus.reached \land consensus.conf \geq \tau$}{
        \tcp{Skip additional debate rounds if consensus reached}
        \textbf{goto} Phase 4\;
    }
    
    \tcp{Phase 3: Multi-round Debate (Rounds 2+)}
    \For{$round = 2$ \KwTo $R_{max}$}{
        $context \leftarrow$ \AR{$responses$}\;
        \ForEach{$a_i \in A$}{
            $prompt \leftarrow$ \GDP{$a_i, context$}\;
            $responses[a_i] \leftarrow$ \QA{$a_i, prompt$}\;
        }
        $consensus \leftarrow$ \ME{$responses, round$}\;
        \If{$consensus.reached \land consensus.conf \geq \tau$}{
            \textbf{break}\;
        }
    }
    
    \tcp{Phase 4: Final Judgment (Judge always decides)}
    $judge\_prompt \leftarrow$ \GJP{$responses, round$}\;
    $decision \leftarrow$ \QJ{$judge\_prompt$}\;
    \Return{$(decision.assess, decision.conf, round, consensus.reached)$}\;
}
\end{algorithm}

\subsection{Specialist Agents}

The PhishDebate framework employs four specialized agents, each designed to analyze specific aspects of phishing detection. Each agent operates with domain-specific prompts and contributes unique insights to the collaborative decision-making process.

\subsubsection{URL Analyst Agent}

The URL Analyst Agent specializes in examining URL structures and domain characteristics to identify phishing indicators. This agent focuses on detecting suspicious patterns in domain names, subdomains, URL paths, and overall URL construction that are commonly associated with phishing attacks. The Fig.~\ref{fig:url_prompt} details the prompt template used for URL Analyst Agent.

\begin{figure}[!ht]
\centering
\begin{minipage}{\linewidth}
\footnotesize
\begin{tcolorbox}[
    colback=white,
    colframe=black,
    coltitle=white,
    colbacktitle=black,
    fonttitle=\bfseries\ttfamily,
    title=Prompt Template 1: URL Analyst Agent,
    width=\linewidth,
    boxrule=0.5mm,
    left=2mm, right=2mm, top=1mm, bottom=1mm
]
\ttfamily
\textbf{Input:} Website URL ($U$)

\vspace{0.5em}
\textbf{Prompt Template:}
\begin{quote}
\textit{You are a cybersecurity expert specializing in URL analysis for phishing detection. Examine the provided URL and identify suspicious patterns, domain characteristics, subdomain usage, URL structure, and any indicators that suggest phishing or legitimate intent.}

\textit{URL: [TARGET\_URL]}

\textit{Provide your response in the following format:} \\
\textit{- Claim: [Your phishing/non-phishing assessment of the URL]} \\
\textit{- Confidence: [A score between 0 and 1]} \\
\textit{- Evidence: [Key suspicious or benign patterns you found]}
\end{quote}
\end{tcolorbox}
\end{minipage}
\caption{Prompt template for URL Analyst Agent.}
\label{fig:url_prompt}
\end{figure}

With this, a structured assessment can be returned containing claim (PHISHING/LEGITIMATE), confidence score (0-1), and supporting evidence focusing on URL-based indicators such as domain spoofing, suspicious TLDs, URL shortening, and deceptive subdomain usage.

\subsubsection{HTML Structure Agent}

The HTML Structure Agent analyzes the underlying HTML code structure to identify technical indicators of phishing websites. This agent examines form elements, JavaScript usage, iframe implementations, and other structural patterns that may indicate malicious intent. The Fig.~\ref{fig:html_prompt} details the prompt template used for HTML Structure Agent.

\begin{figure}[!ht]
\centering
\begin{minipage}{\linewidth}
\footnotesize % ↓↓ Reduce font size from here ↓↓
\begin{tcolorbox}[
    colback=white,
    colframe=black,
    coltitle=white,
    colbacktitle=black,
    fonttitle=\bfseries\ttfamily,
    title=Prompt Template 2: HTML Structure Agent,
    width=\linewidth,
    boxrule=0.5mm,
    left=2mm, right=2mm, top=1mm, bottom=1mm
]
\ttfamily
\textbf{Input:} HTML content ($H$)

\vspace{0.5em}
\textbf{Prompt Template:}
\begin{quote}
\textit{You are an expert in web security. Review the HTML structure of a webpage and determine if it exhibits suspicious structural characteristics typical of phishing sites. Focus on elements such as hidden forms, suspicious input fields, iframe usage, obfuscated JavaScript, and deceptive redirection patterns.}

\textit{HTML: [TRUNCATED\_HTML\_CONTENT]}

\textit{Provide your response in the following format:} \\
\textit{- Claim: [Your assessment about the HTML structure indicating phishing or not]} \\
\textit{- Confidence: [A score between 0 and 1]} \\
\textit{- Evidence: [Relevant structural elements or tag patterns you found]}
\end{quote}
\end{tcolorbox}
\end{minipage}
\caption{Prompt template for HTML Structure Agent.}
\label{fig:html_prompt}
\end{figure}

As a result, a technical assessment can be formed focusing on HTML structural anomalies, suspicious form actions, hidden elements, JavaScript obfuscation, and other code-level indicators of phishing attempts.

\subsubsection{Content Semantic Agent}

The Content Semantic Agent performs natural language analysis of the visible website content to detect linguistic patterns and semantic cues associated with phishing attacks. This agent specializes in identifying manipulative language, urgency tactics, and social engineering techniques. The Fig.~\ref{fig:content_prompt} illustrates the prompt template used for Content Semantic Agent.

\begin{figure}[!ht]
\centering
\begin{minipage}{\linewidth}
\footnotesize % ↓ Apply reduced font size from here ↓
\begin{tcolorbox}[
    colback=white,
    colframe=black,
    coltitle=white,
    colbacktitle=black,
    fonttitle=\bfseries\ttfamily,
    title=Prompt Template 3: Content Semantic Agent,
    width=\linewidth,
    boxrule=0.5mm,
    left=2mm, right=2mm, top=1mm, bottom=1mm
]
\ttfamily
\textbf{Input:} Visible text content ($T$)

\vspace{0.5em}
\textbf{Prompt Template:}
\begin{quote}
\textit{You are a cybersecurity-focused language expert. Read the visible text content extracted from a webpage and decide whether the language indicates phishing intent. Look for emotionally manipulative language, requests for sensitive information, login instructions, urgency, or impersonation of known organizations.}

\textit{Visible Text: [TRUNCATED\_VISIBLE\_TEXT]}

\textit{Provide your response in the following format:} \\
\textit{- Claim: [Whether the page language seems phishing-related]} \\
\textit{- Confidence: [A score between 0 and 1]} \\
\textit{- Evidence: [Specific words, phrases, or sentence patterns that support your claim]}
\end{quote}
\end{tcolorbox}
\end{minipage}
\caption{Prompt template for Content Semantic Agent.}
\label{fig:content_prompt}
\end{figure}

Through this linguistic analysis, emotional manipulation, urgency indicators, credential harvesting language, and other semantic patterns characteristic of phishing content can be identified.

\subsubsection{Brand Impersonation Agent}

The Brand Impersonation Agent specializes in detecting attempts to impersonate legitimate brands or organizations. This agent analyzes both URL and content elements to identify unauthorized use of brand names, logos, and corporate identity elements. The Fig.~\ref{fig:brand_prompt} demonstrates the prompt template used for Brand Impersonation Agent.

\begin{figure}[!ht]
\centering
\begin{minipage}{\linewidth}
\footnotesize % ↓ Apply reduced font size from here ↓
\begin{tcolorbox}[
    colback=white,
    colframe=black,
    coltitle=white,
    colbacktitle=black,
    fonttitle=\bfseries\ttfamily,
    title=Prompt Template 4: Brand Impersonation Agent,
    width=\linewidth,
    boxrule=0.5mm,
    left=2mm, right=2mm, top=1mm, bottom=1mm
]
\ttfamily
\textbf{Input:} URL ($U$) and visible text content ($T$)

\vspace{0.5em}
\textbf{Prompt Template:}
\begin{quote}
\textit{You are a brand impersonation detection expert. Based on the URL and the HTML-visible content, evaluate whether this page attempts to impersonate a known brand. Focus on brand names, company references, login language, and any indications of misused identity (such as pretending to be Google, Apple, PayPal, etc.).}

\textit{URL: [TARGET\_URL]} \\
\textit{Visible Text:}\\
\textit{[TRUNCATED\_VISIBLE\_TEXT]}

\textit{Provide response in following format:} \\
\textit{- Claim: [Does the content attempt to impersonate a known brand?]} \\
\textit{- Confidence: [A score between 0 and 1]} \\
\textit{- Evidence: [Name(s) of impersonated brands and supporting context]}
\end{quote}
\end{tcolorbox}
\end{minipage}
\caption{Prompt template for Brand Impersonation Agent.}
\label{fig:brand_prompt}
\end{figure}

The brand impersonation assessment identifies the specific brands being mimicked, authenticity indicators, and evidence of legitimate versus fraudulent brand usage.

\subsection{Coordination Agents}

\subsubsection{Moderator}

The Moderator serves as the debate coordinator, evaluating specialist agent responses after each round to determine if consensus has been reached. The Moderator analyzes the collective insights from all active specialist agents and decides whether the evidence is sufficient for a confident determination. The prompt template used for Moderator can be found in Appendix~\ref{app:prompts} Fig.~\ref{fig:moderator_prompt}.

A JSON-formatted consensus evaluation will be generated containing: consensus status (Yes/No), supported assessment (PHISHING/LEGITIMATE/UNCERTAIN), detailed reasoning, confidence score, and continuation decision.

\subsubsection{Judge}

The Judge makes the final determination when no consensus is reached after the maximum number of debate rounds. The Judge considers all evidence presented by specialist agents throughout the entire debate process and renders a definitive verdict. The prompt template used for Judge can be found in Appendix~\ref{app:prompts} Fig.~\ref{fig:final_judgment_prompt}.

The Judge Agent outputs the Final judgment in JSON format containing definitive assessment (PHISHING/LEGITIMATE), confidence score, comprehensive reasoning, and key evidence summary.

PhishDebate follows a modular design philosophy that enables flexible configuration based on specific use cases and requirements. The framework supports several customization options:

\textbf{Agent Exclusion:} Users can selectively exclude specific specialist agents from the debate process while maintaining system integrity. The framework enforces safety constraints ensuring that critical agents (Moderator and Judge) cannot be excluded and at least one specialist agent remains active. This feature enables comparative analysis and performance optimization for different scenarios.

\textbf{Configurable Debate Rounds:} The number of debate rounds can be adjusted (1-10 rounds) to balance between thoroughness and computational efficiency. Early termination mechanisms allow the system to conclude debates when strong consensus is reached, optimizing resource utilization.

In addition, to optimize token usage and prevent context limit violations, PhishDebate implements intelligent content truncation for HTML and text inputs. HTML content is truncated based on model-specific token limits, with truncation performed at HTML tag boundaries to preserve structural integrity. A truncation notice is appended when content exceeds limits, ensuring agents are aware of incomplete data while maintaining computational efficiency.

\section{Evaluation}\label{sec:eval}

\subsection{Models Used}
To evaluate the performance of the proposed \textit{PhishDebate} framework, we selected four commercial LLMs for comparison and further analysis: Qwen2.5-vl-72b-instruct\cite{bai2023qwen}, Gemini-2.0-Flash\cite{google2023gemini}, GPT-4o\cite{openai2023gpt4}, and GPT-4o Mini\cite{openai2023gpt4}. These models were chosen for their advanced reasoning capabilities, diverse architectures, and wide availability via Application Programming Interface (API), providing a representative benchmark across different commercial LLM ecosystems.

\subsection{Datasets \& data Processing}
This study employs two datasets to evaluate and analyze the effectiveness of PhishDebate. The first is a Phishing Websites Dataset from Mendeley \cite{MendeleyPhishingDataset}, consisting of both phishing and legitimate websites. This dataset originally includes URL, HTML source, and metadata, and was collected from various sources including Google Search, Ebbu2017 Phishing Dataset\cite{Ebbu2017}, PhishTank\cite{PhishTank}, OpenPhish\cite{OpenPhish}, and PhishRepo\cite{PhishRepo}. For our experiments, we randomly sampled 500 phishing and 500 legitimate instances.

The second dataset is the TR-OP Dataset\cite{li2024knowphish}, which contains manually labeled and balanced samples. Benign samples are drawn from the Tranco \cite{LePochat2019Tranco} top 50k domains, while phishing samples were crawled and validated within a six-month period from July to December 2023, spanning 440 unique phishing targets from OpenPhish\cite{OpenPhish}. Similarly, we randomly selected 500 phishing and 500 legitimate samples from this dataset.

We used the Mendeley dataset for performance benchmarking and comparison across models, while both Mendeley and TR-OP datasets were used in the scenario analysis to enhance result robustness.

For data processing, we implemented a pipeline to extract structured features from each sample. The procedure includes: (1) reading and verifying the URL and raw HTML; (2) using BeautifulSoup to parse and clean HTML by removing tags like \texttt{style}, \texttt{noscript}, and external stylesheets; (3) extracting visible text by removing all \texttt{script} tags. This produces three components per sample: URL, cleaned HTML, and visible text.

The following Algorithm \ref{alg:html_preprocess} summarizes our data preprocessing pipeline:

\begin{algorithm}
\caption{Data Preprocessing Pipeline}
\label{alg:html_preprocess}
\footnotesize
\KwIn{Sample $s = (f_U, f_H)$ where $f_U$: URL file, $f_H$: HTML file}
\KwOut{Processed tuple $(U, H, T)$: URL, cleaned HTML, and visible text}

$U \leftarrow$ Read($f_U$), $R \leftarrow$ Read($f_H$) \tcp*{Extract URL and raw HTML}
\If{$U = \emptyset$ or $R = \emptyset$}{\Return{\texttt{None}} \tcp*{Skip invalid sample}}
$H' \leftarrow$ CleanHTML($R$) \tcp*{Remove \texttt{style}, \texttt{noscript}, \texttt{link[rel=stylesheet]}}
$H \leftarrow$ Remove($H'$, \texttt{script}) \tcp*{Final cleaned HTML}
$T \leftarrow$ ExtractText($H$) \tcp*{Visible text from HTML}
\Return{$(U, H, T)$}
\end{algorithm}

\subsection{Evaluation Metrics}
We evaluate model performance using standard binary classification metrics, computed from the confusion matrix:

\begin{itemize}
\item \textbf{True Positive Rate (TPR)} / \textbf{Recall}: $\text{TPR} = \frac{TP}{TP + FN}$ \ Measures the proportion of phishing samples correctly identified.

\item \textbf{True Negative Rate (TNR)}: $\text{TNR} = \frac{TN}{TN + FP}$ \ 
Measures the proportion of legitimate samples correctly identified.

\item \textbf{False Positive Rate (FPR)}: $\text{FPR} = \frac{FP}{FP + TN}$ \ 
Measures how often legitimate websites are incorrectly classified as phishing.

\item \textbf{False Negative Rate (FNR)}: $\text{FNR} = \frac{FN}{FN + TP}$ \ 
Measures how often phishing websites are missed.

\item \textbf{Precision}: $\text{Precision} = \frac{TP}{TP + FP}$ \ 
Measures the correctness of phishing predictions.

\item \textbf{Accuracy}: $\text{Accuracy} = \frac{TP + TN}{TP + TN + FP + FN}$ \ 
Reflects the overall classification correctness.

\item \textbf{F1 Score}: $\text{F1} = 2 \cdot \frac{\text{Precision} \cdot \text{Recall}}{\text{Precision} + \text{Recall}}$ \ 
Harmonic mean of precision and recall; balances false positives and false negatives.
\end{itemize}

\section{Results}\label{sec:results}
\subsection{Performance Evaluation}
Table~\ref{tab:llm_eval_metrics} illustrates the results of performance evaluation across four commercial LLMs on the PhishDebate framework, including GPT-4o, GPT-4o-mini, Gemini-2.0, and Qwen2.5-vl-72b-instruct. The results demonstrate that GPT-4o consistently outperforms others in terms of overall classification effectiveness, achieving the highest accuracy (96.50\%), precision (94.97\%), and F1 score (96.56\%), along with the best true negative rate (94.8\%) and the lowest false positive rate (5.2\%). Gemini-2.0, while exhibiting the fastest inference time (17.9 s), achieved the highest true positive rate (99.2\%) and the lowest false negative rate (0.8\%), indicating strong phishing detection capability. However, its performance was affected by 8 instances in which it failed to follow the binary classification instruction in the prompt, instead outputting an uncertain response. These cases, originating from legitimate websites, were treated as misclassifications, thereby increasing Gemini's false positive rate to 14.4\%, lowering its true negative rate to 85.57\%, and reducing its overall accuracy and F1 score to 92.44\% and 92.97\%, respectively. GPT-4o-mini performs competitively but lags slightly behind GPT-4o, whereas Qwen showed the lowest accuracy and F1 score due to a significantly higher false negative rate (19.4\%), despite maintaining a relatively low false positive rate (9.4\%). These results highlight GPT-4o as the most balanced and reliable model for phishing detection in the context of the PhishDebate framework.

\renewcommand{\arraystretch}{1.3} % increase line spacing in the table
\begin{table*}[!h]
\caption{Evaluation Metrics of PhishDebate on Commercial LLMs}
\centering
\begin{tabular}{lccccccccc}
\toprule
Model & TPR & TNR & FPR & FNR & Recall & Precision & Accuracy & F1 Score & Time (s) \\
\midrule
\textbf{GPT-4o} & 0.982 & \textbf{0.948} & \textbf{0.052} & 0.018 & 0.982 & \textbf{0.9497} & \textbf{0.9650} & \textbf{0.9656} & 22.20 \\
GPT-4o-mini & 0.980 & 0.898 & 0.102 & 0.020 & 0.980 & 0.9057 & 0.9390 & 0.9414 & 37.50 \\
Gemini-2.0-Flash & \textbf{0.992} & 0.8557 & 0.1443 & \textbf{0.008} & \textbf{0.992} & 0.8748 & 0.9244 & 0.9297 & \textbf{17.90} \\
Qwen2.5-vl-72b-instruct & 0.806 & 0.906 & 0.094 & 0.194 & 0.806 & 0.8956 & 0.8560 & 0.8484 & 56.36 \\
\bottomrule
\end{tabular}
\label{tab:llm_eval_metrics}
\end{table*}

\subsection{Comparative Evaluation}
To evaluate the effectiveness of our proposed multi-agent debate system, we compare it against two widely adopted single-agent prompting strategies. First, we include a Single-Agent with Direct Prompt baseline, where an LLM receives the raw input (URL and HTML-visible text) and directly predicts whether the page is phishing without any intermediate reasoning. This baseline reflects a straightforward application of LLMs for classification tasks and serves as a reference point for measuring improvements attributable to structured reasoning. Second, we include a Single-Agent with CoT baseline \cite{NEURIPS2022_9d560961}, where the LLM is prompted to reason step-by-step through URL analysis, content interpretation, and brand impersonation detection before arriving at a decision. This variant captures the benefits of explicit intermediate reasoning within a single model, enabling us to isolate and assess the added value of distributed agent roles and debate-based aggregation in our system. We provide the prompt templates used for two baselines in Appendix~\ref{app:prompts} Fig.~\ref{fig:direct_prompt} and Fig.~\ref{fig:cot_prompt}.

Table \ref{tab:baseline_comparison} presents the results of comparative evaluation between the proposed PhishDebate framework and two baselines which are Single Agent and CoT prompting. The results demonstrate the effectiveness of our debate-driven approach in enhancing phishing detection on GPT-4o-mini. PhishDebate outperforms both baselines across all key evaluation metrics, achieving the highest precision (90.57\%), accuracy (93.90\%), recall (98.00\%), and F1 score (94.14\%). Notably, while CoT prompting provides reasonably strong results (e.g., 90.94\% F1 score), it suffers from lower recall and precision compared to PhishDebate, especially when factoring in the 50 cases it labeled as “uncertain” instead of providing definitive classifications. These include 16 phishing websites and 34 legitimate websites, which, according to our evaluation protocol, are considered misclassifications. This underscores a critical limitation of CoT in reliability and confidence. In contrast, the PhishDebate framework significantly improves the confidence and decisiveness of LLM-based predictions by leveraging multi-agent debate, thereby reducing indecisive outputs and promoting consistent judgment across phishing and legitimate classes.

In terms of efficiency, the three approaches exhibit clear differences in average inference time (Table~\ref{tab:baseline_comparison}). The Single Agent baseline is the fastest, requiring only 4.7 s per sample. CoT increases the processing cost to 10.5 s. By contrast, PhishDebate achieves the strongest detection effectiveness but requires 37.5 s. Despite the higher latency and additional tokens consumed by multi-round debates, the overall cost of running PhishDebate remains modest: processing 1,000 samples incurred only about \$3.357 in inference expenses.

\begin{table}[!t]
\caption{Comparison of PhishDebate with Single Agent and CoT}
\centering
\renewcommand{\arraystretch}{1.2}
\setlength{\tabcolsep}{4pt}
\begin{tabular}{lccccc}
\toprule
Method & Prec. & Acc. & Rec. & F1 & Time \\
\midrule
Single Agent & 0.6057 & 0.6700 & 0.9740 & 0.7469 & \textbf{4.7 s↓} \\
CoT          & 0.8861 & 0.9070 & 0.9340 & 0.9094 & 10.5 s \\
\textbf{PhishDebate} & \textbf{0.9057↑} & \textbf{0.9390↑} & \textbf{0.9800↑} & \textbf{0.9414↑} & 37.5 s \\
\bottomrule
\end{tabular}
\footnotesize \textit{Note.} Prec. = Precision; Acc. = Accuracy; Rec. = Recall; F1 = F1 Score; Time = Avg. Time (seconds).
\label{tab:baseline_comparison}
\end{table}

\subsection{Scenario Analysis}
PhishDebate follows a modular design, allowing users to flexibly exclude specific specialized agents according to different use case requirements, such as resource constraints or application priorities.

Table~\ref{tab:agent_exclusion_analysis} presents a scenario analysis evaluating the performance of the PhishDebate framework when individual specialized agents are excluded. The results are averaged across two datasets (Mendeley and TR-OP) to ensure robustness. The full configuration using all agents achieves strong overall performance, particularly demonstrating the highest recall (0.985) and lowest false negative rate (FN = 8), indicating its superior ability to identify phishing websites with minimal missed detections. This highlights the importance of the ensemble debate mechanism in capturing diverse phishing indicators across URL, HTML, semantic, and brand features.

Notably, the exclusion of the HTML Structure Agent leads to the highest F1 score (0.9471), accuracy (0.9455), precision (0.9207), and the lowest false positive rate (FP = 42), suggesting that in specific deployment scenarios where legitimate website preservation is critical, a reduced configuration excluding this agent may offer better precision-legitimacy trade-offs. However, this comes at a minor cost in recall (0.975), which is still lower than the full configuration.

The variant without the Content Semantic Agent matches the full agent setup in recall and true positive rate (TP = 492), but suffers in terms of increased false positives (FP = 56), suggesting a reduced ability to discriminate legitimate content. This supports the value of semantic reasoning in reducing over-flagging. Similarly, excluding the URL Analyst Agent leads to a clear degradation in phishing detection capability, with a recall drop to 0.952 and increased false negatives (FN = 24), reinforcing the foundational role of URL features in phishing identification. Finally, omitting the Brand Impersonation Agent results in slightly lower overall performance, underscoring its contribution to brand spoof detection. Overall, these findings emphasize the effectiveness of the modular PhishDebate design, while demonstrating that task-specific configurations can still perform competitively under constrained scenarios.

\begin{table*}[!h]
\renewcommand{\arraystretch}{1.3}
\caption{Scenario Analysis of Agent Exclusion in PhishDebate Framework}
\centering
\begin{tabular}{lrrrrrrrrr}
\toprule
         Setting &  Avg. TP &  Avg. FN &  Avg. FP &  Avg. TN &  Avg. Recall &  Avg. Precision &  Avg. Accuracy &  Avg. F1 Score \\
\midrule
All Agents         & \textbf{492} & \textbf{8}  &  50 & 450 & \textbf{0.985} & 0.9070 & 0.9420 & 0.9444 \\
W/O URL Agent               & 476 &  24 &  46 & 454 & 0.952 & 0.9110 & 0.9295 & 0.9311 \\
W/O HTML Agent     & 488 &  12 & \textbf{42} & \textbf{458} & 0.975 & \textbf{0.9207} & \textbf{0.9455} & \textbf{0.9471} \\
W/O Content Agent  & \textbf{492} & \textbf{8}  &  56 & 444 & \textbf{0.985} & 0.8987 & 0.9370 & 0.9399 \\
W/O Brand Agent             & 489 &  11 &  56 & 444 & 0.978 & 0.8981 & 0.9335 & 0.9363 \\
\bottomrule
\end{tabular}
\label{tab:agent_exclusion_analysis}
\end{table*}

\subsection{Case Study}
PhishDebate begins each investigation by distributing the same artefacts including the raw URL\footnote{The raw URL for this phishing website is: \url{https://mail.mxcapital.com.br/wp-includes/wells/wells/page/index.htm}.}, rendered HTML, and visible text to its four specialised agents (URL Analyst, HTML Structure, Content Semantic, and Brand-Impersonation). Each agent reasons independently within its expertise, returns a \textit{claim} (phishing or legitimate), a numerical \textit{confidence}, and traceable \textit{evidence}. The moderator then inspects the four perspectives and decides whether a clear majority exists. If not, it compels another debate round, asking every agent to reconcile its view with the counter-evidence provided by its peers. Only when consensus is reached (or the maximum number of rounds elapses) does the moderator submit a verdict to the judge, who finalises the assessment and records a confidence score.

\begin{figure}[!t]
\centering
\includegraphics[width=\linewidth]{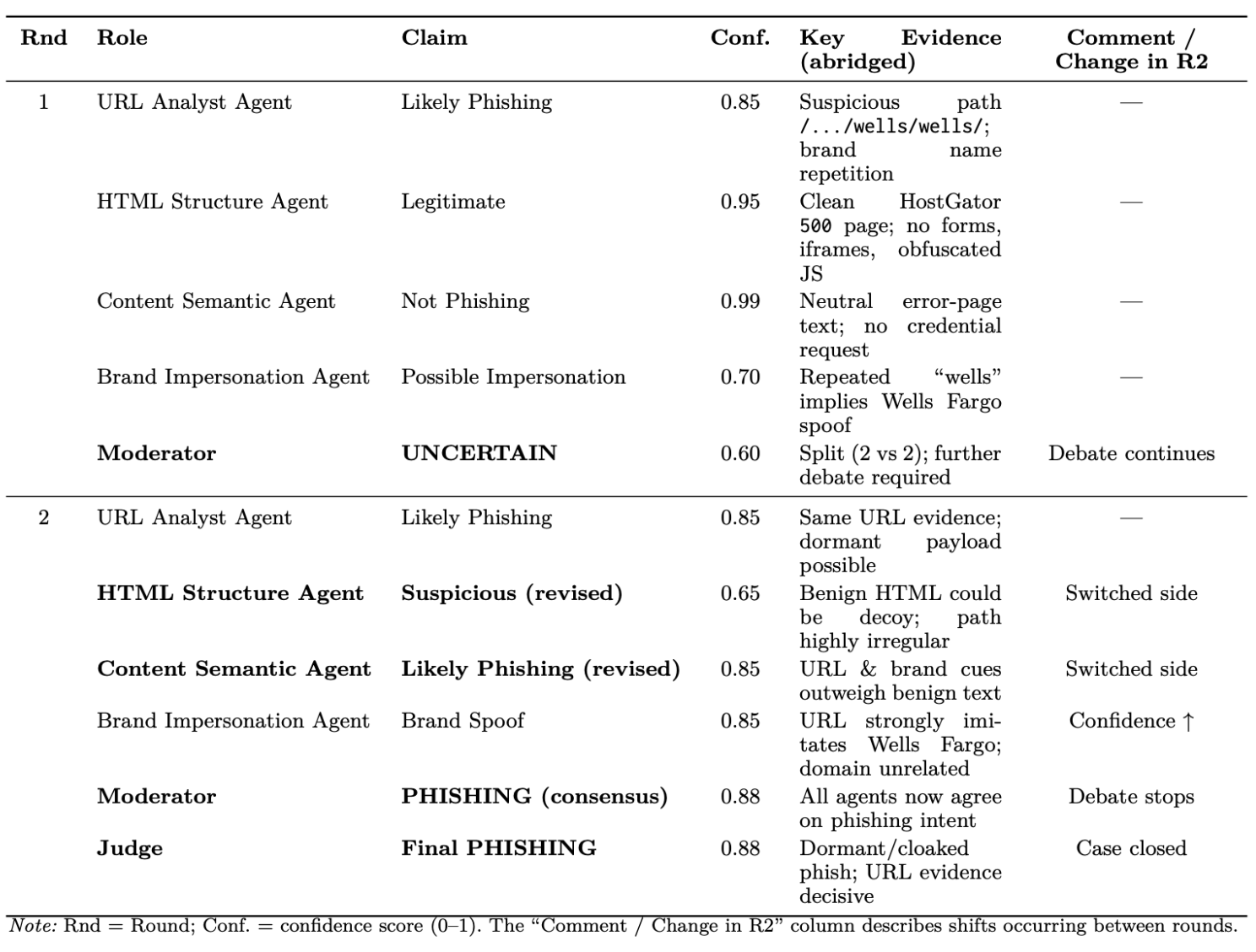}
\caption{A detailed detection example with PhishDebate.}
\label{fig:casestudy}
\end{figure}

Fig. \ref{fig:casestudy} narrates a representative two-round session on a suspicious phishing webpage. In Round 1 the agents split 2–2: the URL Analyst and Brand-Impersonation agents flagged the directory string “\texttt{/wp-includes/wells/wells/}” as brand spoofing (Wells Fargo) and noted that legitimate sites never expose \texttt{/wp-includes/}; by contrast, the HTML Structure and Content Semantic agents found no typical phishing artefactssuch as forms, iframes, or persuasive language, and therefore judged the page legitimate. Lacking consensus, the moderator declared the case \textit{UNCERTAIN} and requested a second round.

During Round 2 each agent re-examined the shared evidence. Crucially, the HTML Structure and Content Semantic agents acknowledged that benign error pages are frequently used as decoys and that the highly irregular path outweighed the absence of active credential-stealing code. After switching their stance to “Likely Phishing,” all four agents converged. The moderator therefore recorded a consensus of \textit{PHISHING} with 0.88 confidence, and the judge confirmed the decision. This outcome underscores two salient strengths of PhishDebate: (i) its multi-view debate can uncover \emph{dormant} or \emph{cloaked} phishing infrastructure that content-only scanners miss, and (ii) the step-wise trace in Fig. \ref{fig:casestudy} provides transparent justifications that facilitate analyst review or automated policy triggers.

\section{Limitations}\label{sec:limitations}
Despite its effectiveness, several limitations remain. First, the performance of debate agents ultimately reflects the biases and knowledge gaps of the underlying LLMs, which are bounded by their pre-training data. Second, input length restrictions of commercial LLM APIs constrain the amount of webpage content that can be fully processed, leading to truncation and potential information loss. Third, although our experiments employed commercial APIs, the core contribution of PhishDebate lies in demonstrating how debate-based reasoning mitigates hallucination and enhances interpretability. In practice, the same paradigm can be instantiated with locally deployed language models, enabling large-scale deployment under high-traffic conditions while simultaneously reducing cost and latency.

\section{Conclusion and Future Directions}\label{sec:conclusion}
This paper introduced PhishDebate, a modular multi-agent LLM-based debate framework for phishing website detection. By leveraging collaborative reasoning among specialized agents under the coordination of a Moderator and a final Judge, PhishDebate achieved strong detection performance and addressed key limitations of traditional and single-agent LLM approaches, including limited interpretability and vulnerability to hallucination.  

Future research will examine the resilience of PhishDebate against adversarial phishing tactics, such as obfuscation, cloaking, and adversarially crafted webpages, to further validate its adaptability in dynamic and hostile threat environments. Ultimately, we envision PhishDebate as an analyst-augmentation system that not only strengthens early, left-of-exploit detection but also scales to high-velocity, large-scale security data environments.

\bibliographystyle{IEEEtran}
\bibliography{ref}

\clearpage
\appendices
\section{Prompt Templates Used in This Study}\label{app:prompts}

\begin{figure}[!ht]
\centering
\begin{minipage}{\linewidth}
\scriptsize % ↓ Apply reduced font size from here ↓
\begin{tcolorbox}[
    colback=white,
    colframe=black,
    coltitle=white,
    colbacktitle=black,
    fonttitle=\bfseries\ttfamily,
    title=Prompt Template 5: Moderator Evaluation,
    width=\linewidth,
    boxrule=0.5mm,
    left=2mm, right=2mm, top=1mm, bottom=1mm
]
\ttfamily
\textbf{Input:} All specialist agent responses from current round

\vspace{0.5em}
\textbf{Prompt Template:}
\begin{quote}
\textit{You are a moderator overseeing a phishing detection debate between specialist agents. After each round, you will evaluate all agents' analyses and determine if there is a clear consensus for a phishing determination. If so, please summarize your reasons for supporting the consensus and give the final answer that you think is correct, and the debate will conclude. If not, the debate will continue to the next round.}
\end{quote}
\end{tcolorbox}
\end{minipage}
\caption{Prompt template for Moderator Evaluation.}
\label{fig:moderator_prompt}
\end{figure}

\begin{figure}[!ht]
\centering
\begin{minipage}{\linewidth}
\scriptsize % ↓ Apply reduced font size from here ↓
\begin{tcolorbox}[
    colback=white,
    colframe=black,
    coltitle=white,
    colbacktitle=black,
    fonttitle=\bfseries\ttfamily,
    title=Prompt Template 6: Judge Decision,
    width=\linewidth,
    boxrule=0.5mm,
    left=2mm, right=2mm, top=1mm, bottom=1mm
]
\ttfamily
\textbf{Input:} Complete debate history from all specialist agents across all rounds

\vspace{0.5em}
\textbf{Prompt Template:}
\begin{quote}
\textit{You are an expert cybersecurity judge evaluating phishing detection arguments. You will receive analyses from multiple specialized agents and need to make a final determination about whether the website is phishing or legitimate. Consider the evidence from all agents and their confidence levels. You must make a definitive decision between PHISHING or LEGITIMATE.}
\end{quote}
\end{tcolorbox}
\end{minipage}
\caption{Prompt template for Final Judgment.}
\label{fig:final_judgment_prompt}
\end{figure}

\begin{figure}[!ht]
\centering
\begin{minipage}{\linewidth}
\scriptsize % ↓ Apply reduced font size from here ↓
\begin{tcolorbox}[
    colback=white,
    colframe=black,
    coltitle=white,
    colbacktitle=black,
    fonttitle=\bfseries\ttfamily,
    title=Prompt Template 7: Single-Agent Baseline,
    width=\linewidth,
    boxrule=0.5mm,
    left=2mm, right=2mm, top=1mm, bottom=1mm
]
\ttfamily
\textbf{Prompt Template:}
\begin{quote}
\textit{You are an expert cybersecurity analyst specializing in phishing detection. Your task is to analyze websites and determine if they are phishing or legitimate.}

\textit{You will be provided with:} \\
1. URL of the website \\
2. HTML content of the website \\
3. Visible text content extracted from the website

\textit{Your analysis should consider:}
\begin{itemize}
\item URL characteristics (suspicious domains, typosquatting, etc.)
\item HTML structure and content
\item Visual elements and branding
\item Text content and messaging
\item Technical indicators of phishing
\end{itemize}

\textit{You must respond with exactly one of these classifications:}
\begin{itemize}
\item \textbf{PHISHING:} If the website is designed to deceive users or steal information
\item \textbf{LEGITIMATE:} If the website appears to be genuine and trustworthy
\end{itemize}

\textit{Provide your classification followed by a brief explanation of your reasoning.}
\end{quote}
\end{tcolorbox}
\end{minipage}
\caption{Prompt template for Single-Agent Baseline.}
\label{fig:direct_prompt}
\end{figure}

\begin{figure}[!ht]
\centering
\begin{minipage}{0.9\linewidth}
\scriptsize % ↓ Apply reduced font size from here ↓
\begin{tcolorbox}[
    colback=white,
    colframe=black,
    coltitle=white,
    colbacktitle=black,
    fonttitle=\bfseries\ttfamily,
    title=Prompt Template 8: CoT Baseline,
    width=\linewidth,
    boxrule=0.5mm,
    left=2mm, right=2mm, top=1mm, bottom=1mm
]
\ttfamily
\textbf{Prompt Template:}

You are an expert cybersecurity analyst specializing in phishing detection. Your task is to analyze websites and determine if they are phishing or legitimate using a systematic Chain of Thought approach.

You will be provided with: \\
1. URL of the website \\
2. HTML content of the website \\
3. Visible text content extracted from the website

Please analyze the website step-by-step using the following Chain of Thought process:

\textbf{STEP 1: URL ANALYSIS} \\
- Examine the domain name for suspicious patterns \\
- Check for typosquatting (misspellings of legitimate brands) \\
- Look for suspicious TLDs or subdomains \\
- Identify any URL shortening or redirection indicators

\textbf{STEP 2: CONTENT ANALYSIS} \\
- Analyze the HTML structure and quality \\
- Look for suspicious scripts or hidden elements \\
- Check for legitimate branding vs. impersonation attempts \\
- Examine form elements and data collection practices

\textbf{STEP 3: TEXT ANALYSIS} \\
- Review the visible text for urgency tactics \\
- Check for grammar/spelling errors typical of phishing \\
- Look for legitimate contact information \\
- Analyze the overall messaging and tone

\textbf{STEP 4: TECHNICAL INDICATORS} \\
- Check for HTTPS usage and security indicators \\
- Look for suspicious redirects or external links \\
- Examine metadata and technical elements \\
- Consider overall website quality and professionalism

\textbf{STEP 5: FINAL ASSESSMENT} \\
- Weigh all evidence from previous steps \\
- Consider the overall risk profile \\
- Make a final classification with confidence level

\vspace{0.5em}
Format your response as: \\
\texttt{STEP 1: [Your URL analysis]} \\
\texttt{STEP 2: [Your content analysis]} \\
\texttt{STEP 3: [Your text analysis]} \\
\texttt{STEP 4: [Your technical analysis]} \\
\texttt{STEP 5: [Your final assessment]}

\texttt{CLASSIFICATION: [PHISHING or LEGITIMATE]} \\
\texttt{CONFIDENCE: [High/Medium/Low]} \\
\texttt{REASONING: [Brief summary of key factors that led to your decision]}
\end{tcolorbox}
\end{minipage}
\caption{Prompt template for CoT Baseline}
\label{fig:cot_prompt}
\end{figure}

\end{document}